\documentclass{ws-procs975x65}

\def\magstar{{\tt Magstar}}

\def\be {\begin{equation}}
\def\ee {\end{equation}  }
\def\beq{\begin{eqnarray}}
\def\eeq{\end{eqnarray}  }

\begin{document}

\title{Dynamics of  Rotating, Magnetized Neutron Stars}
\author{STEVEN L. LIEBLING}
\address{Physics Department--C.W. Post Campus, Long Island University,\\
Brookville, NY 11548, U.S.A.\\
$^*$E-mail: steve.liebling@liu.edu}

%
%
\begin{abstract}
Using a fully general relativistic implementation of ideal
magnetohydrodynamics with no assumed symmetries in three spatial dimensions,
the dynamics of magnetized, rigidly rotating neutron stars are studied.
Beginning with fully consistent initial data constructed with Magstar,
part of the Lorene project, we study the dynamics and stability of
rotating, magnetized polytropic stars as models of neutron stars.
Evolutions suggest that some of these rotating, magnetized stars
may be minimally unstable occurring at the threshold of black hole
formation.
\end{abstract}


\bodymatter

%
%

\vspace{0.3in}

\noindent{\bf Introduction:~~}
Neutron stars are fascinating astrophysical objects melding
incredibly strong gravity with large densities and pressures.
Along with powerful magnetic fields, these compact stars form
magnetars which are
suspected as the engines behind anomalous X-ray pulsars~(AXPs) and
soft gamma ray repeaters (SGRs)~\cite{Mereghetti:2008je}.
There is a long history of numerical models of neutron stars, but here
their evolution is studied using a code developed for, and applied to, a
number of astrophysical problems~\cite{Palenzuela:2006wp,Palenzuela:2007dm,Anderson:2007kz,Anderson:2008zp,Palenzuela:2009yr,Megevand:2009yx}. This code adopts a fully nonlinear
scheme for the Einstein gravitational field equations which allows for the easy extraction
of gravitational wave signatures. This gravity is coupled to a magnetohydrodynamic~(MHD)
component which uses high resolution shock capturing~(HRSC) methods to evolve a
magnetized fluid in a general relativistic, finite-difference
scheme.
Further details concerning the formulation and numerical methods are presented
in Ref.~\citen{magstar}.

A number of tests are presented in Ref.~\citen{magstar}.
Evolutions of stable, rotating stars demonstrate convergence with increasing resolution.
That is, conservation of quantities such as baryon mass and the $z$-component of the
angular momentum improves as one moves to better resolution. Likewise, violations
of the constraints remain closer to zero with better resolution.

\noindent{\bf Results:~~}
Also discussed in Ref.~\citen{magstar} were evolutions of unstable stellar solutions.
Such solutions either expand, oscillating about some stable star, or
they instead collapse, eventually forming a black hole. Beginning with
a particular unstable solution,
unavoidable numerical error will ultimately drive it to one of these end states.
However, it was found in Ref.~\citen{magstar} that one could tune the perturbation
via some generalized parameter $p$ ({\em e.g.} the amplitude of some perturbation to the initial
pressure) such that for some critical value $p^*$, if $p>p^*$ then the solution collapsed
and if $p<p^*$ the solution expanded. This ability to tune to threshold suggests that the unstable solutions
have just a single unstable mode, and therefore that the solutions represent Type~I critical
solutions.

One question remaining, among many, is whether
initial data far from these unstable solutions in some configuration space
will, upon tuning, evolve towards 
these unstable solutions  and find them at threshold. The work of~Ref.~\citen{Noblephd} answers this question with respect to nonrotating, spherical stars 
in spherical symmetry by evolving
stable TOV solutions perturbed gravitationally by a pulse of scalar field.
For energetic enough scalar pulses, the stable star is sufficiently disturbed that it collapses
to a black hole. Tuning this initial scalar pulse, they find longer and longer lived unstable stellar
solutions.

Because this code already has the capability to evolve a massless, complex scalar field (as studied within the
context of the dynamics of boson stars~\cite{Palenzuela:2006wp,Palenzuela:2007dm}), it is straightforward
to include a scalar pulse as initial data. In particular, the real component of the scalar field $\phi(x,y,z,t)$
is defined as a Gaussian pulse,
$
\phi(t=0)       =  A e^{\left( r - R_0\right)^2/\delta^2}
$,
in terms of a radial coordinate
$
r               \equiv  \sqrt{ \epsilon_x x^2 + \epsilon_y y^2 + z^2}
$ 
with real constants $\epsilon_x$, $\epsilon_y$, $R_0$, $\delta$, and $A$. The imaginary component
remains zero throughout the evolution and we define the initial time derivative of the field such that
$\phi$ is approximately in-going. There are two significant differences with respect to the work of~Ref.~\citen{Noblephd}:
(i) the Einstein field equations are not re-solved to account for the scalar pulse and (ii) the scalar pulse
is located outside the star initially, but not very far from it. In particular, two searches are carried out, both
choosing $R_0=30$ and $\delta=6$, and using $A$ as the generalized parameter $p$ which is tuned to threshold.
The first search uses a spherically symmetric pulse described by  $\epsilon_x=1$, $\epsilon_y=1$, and
the second uses $\epsilon_x=1.25$, $\epsilon_y=0.75$.

\begin{figure}[b]%
\begin{center}
  \parbox{2.1in}{\epsfig{figure=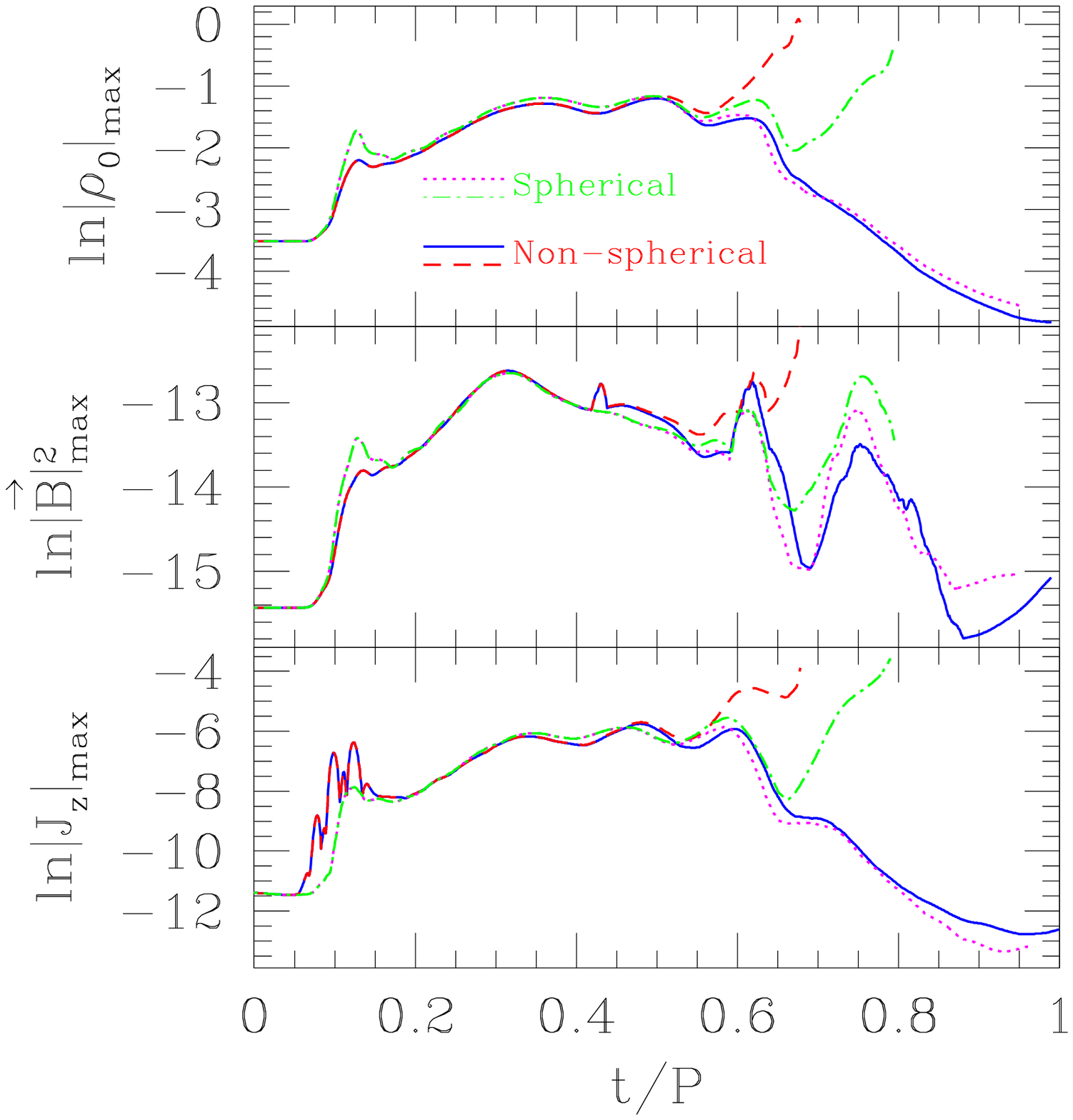,width=2.5in}}
  \hspace*{4pt}
  \parbox{2.1in}{\epsfig{figure=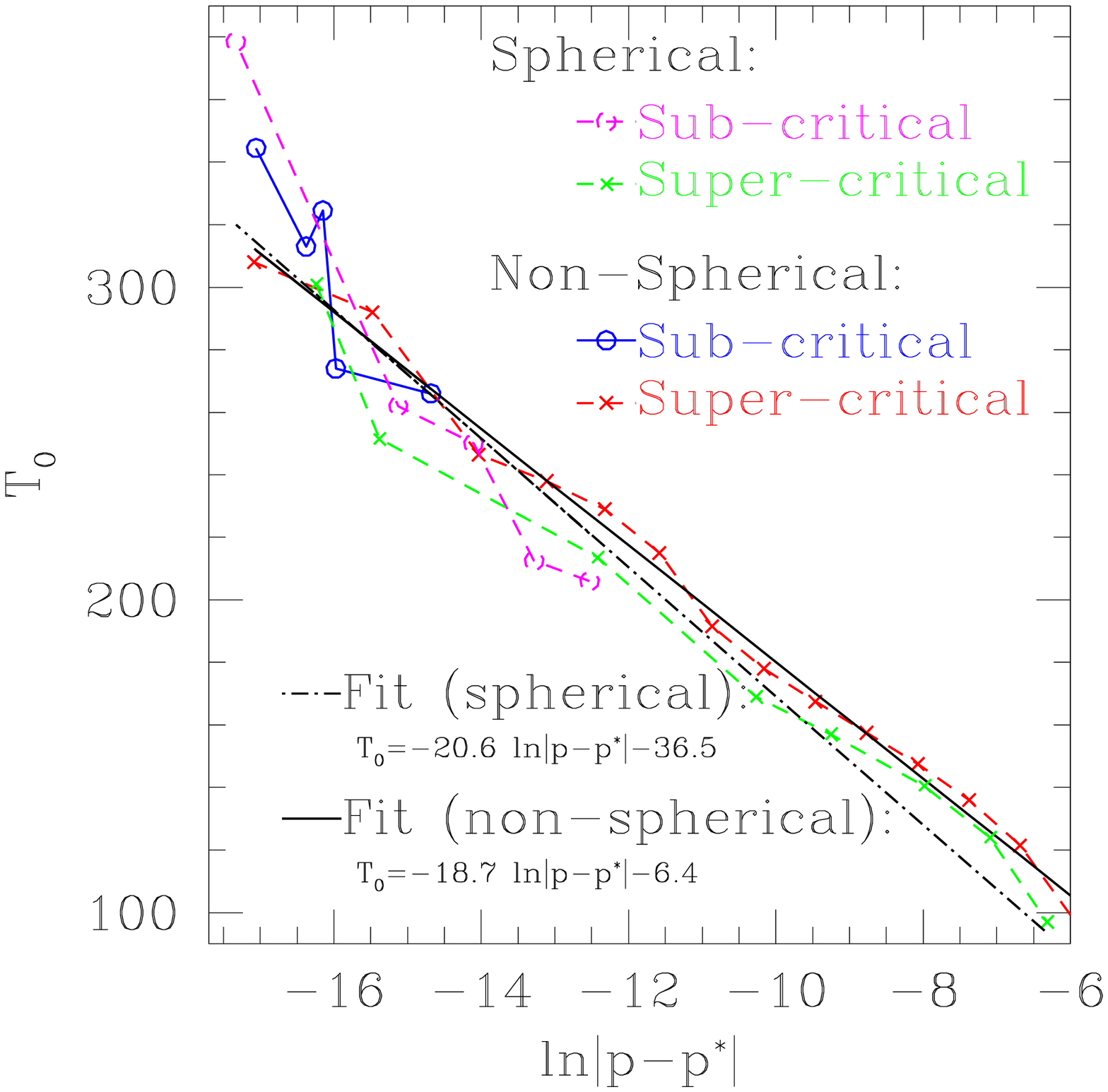,width=2.5in}}
  \caption{Results of two searches using a scalar pulse to perturb the same, magnetic, rotating star.
   {\bf Left:} Stellar properties as functions of time in units of the period of rotation for
   near-critical evolutions.
   {\bf Right:} Survival time of evolutions for the two searches demonstrates the expected scaling
   with the distance from criticality.
  }
  \label{fig:crit}
\end{center}
\end{figure}

It is found that tuning a single parameter
results in the evolution approaching an unstable solution which 
once again suggests that the unstable solution is itself a co-dimension one
unstable solution. The time $T_0$ for which the evolution is near the unstable solution is expected to scale as
$T_0 = -\sigma \ln |p-p^*| + C$ for some constant $C$ that depends on the particular initial-data family and some universal
constant $\sigma$, the inverse of the growth rate of the instability. As indicated by Fig.~\ref{fig:crit}, the results are consistent
with this expected scaling, given the relatively large errors here. These numerical errors  appear to be related
to effects from the boundary which significantly limit how close to criticality one can achieve. These two searches are
both tuned to approximately one part in a million.

\begin{figure}[b]%
\begin{center}
  {\epsfig{figure=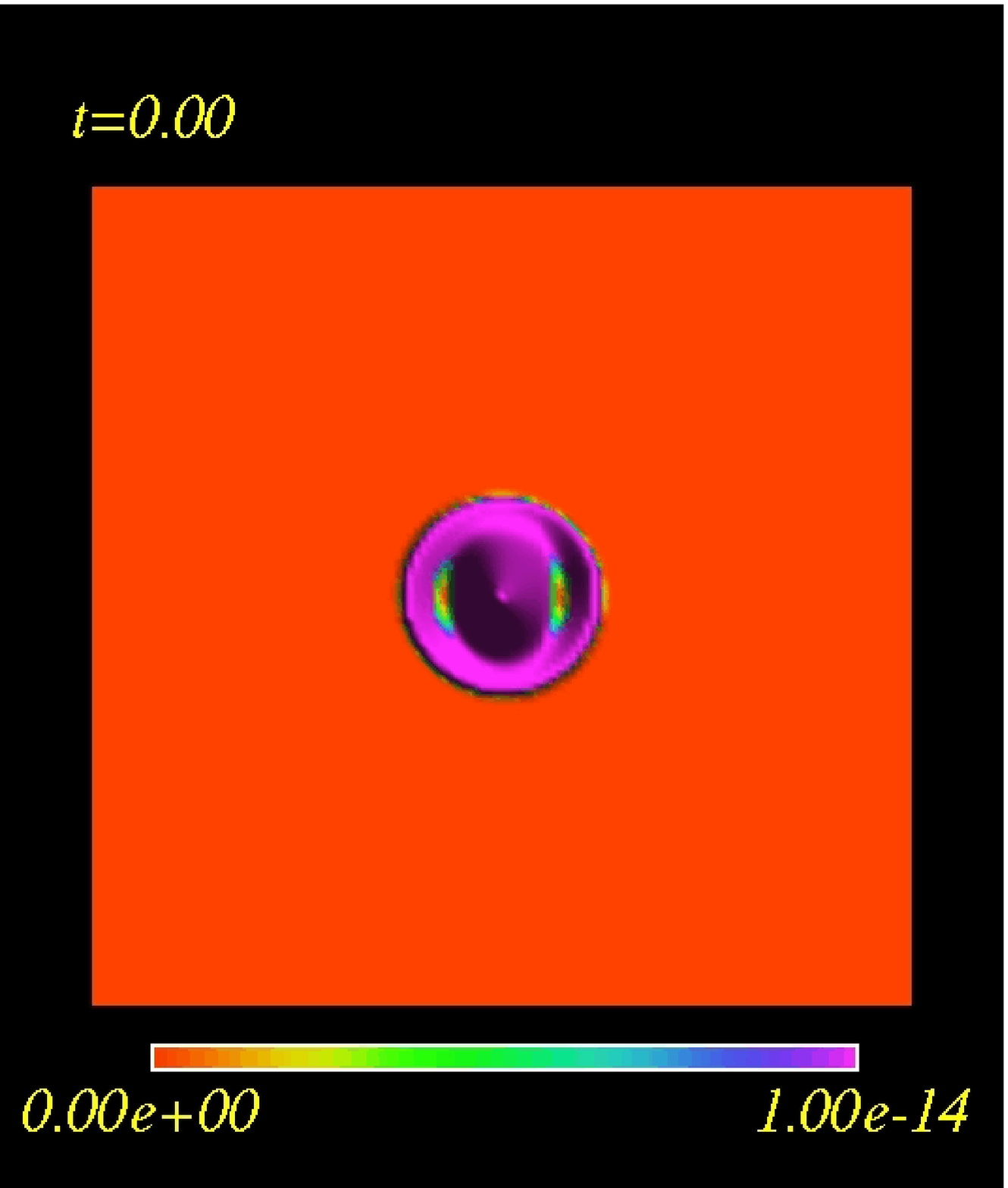,width=1.15in}}
  {\epsfig{figure=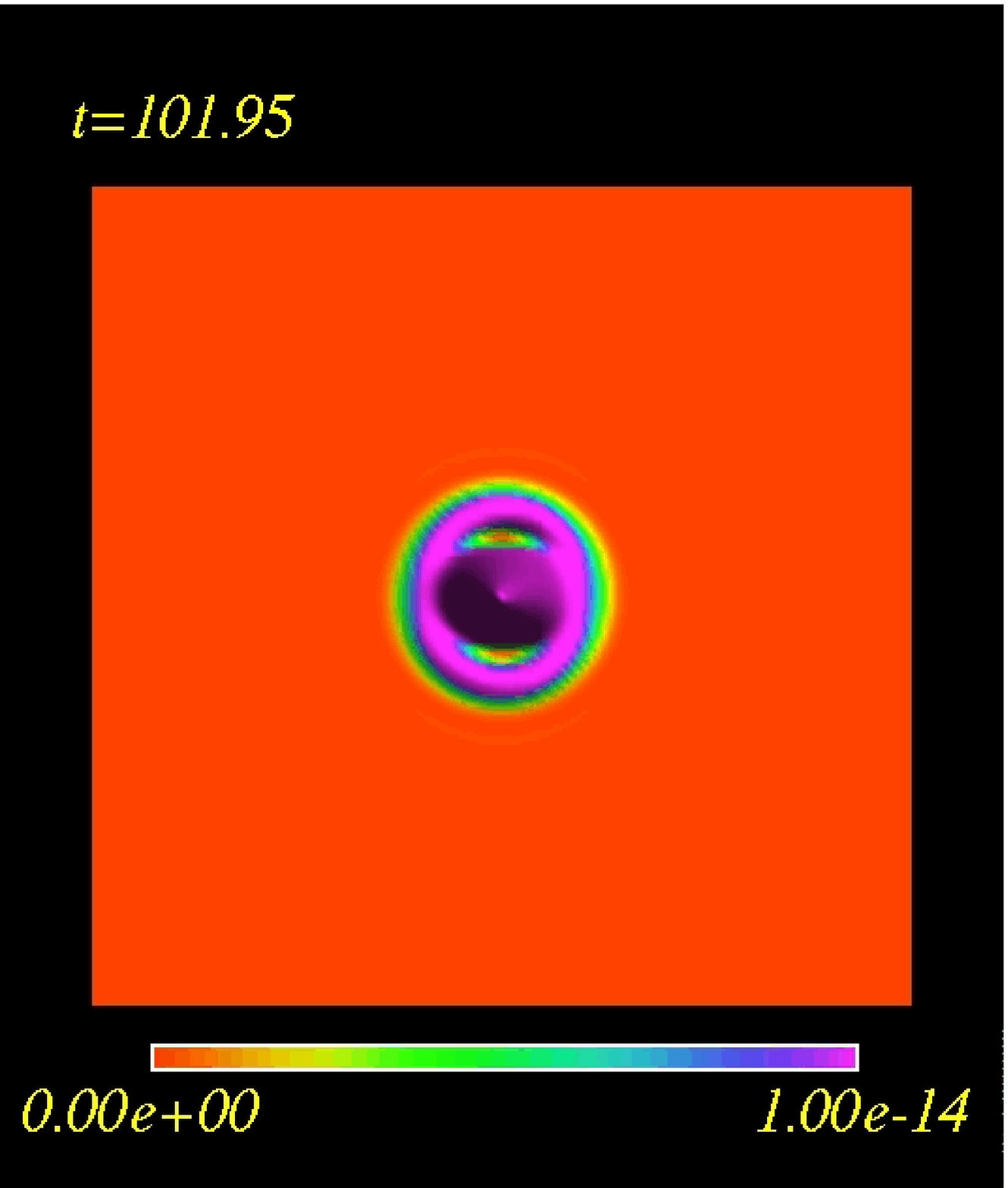,width=1.15in}}
  {\epsfig{figure=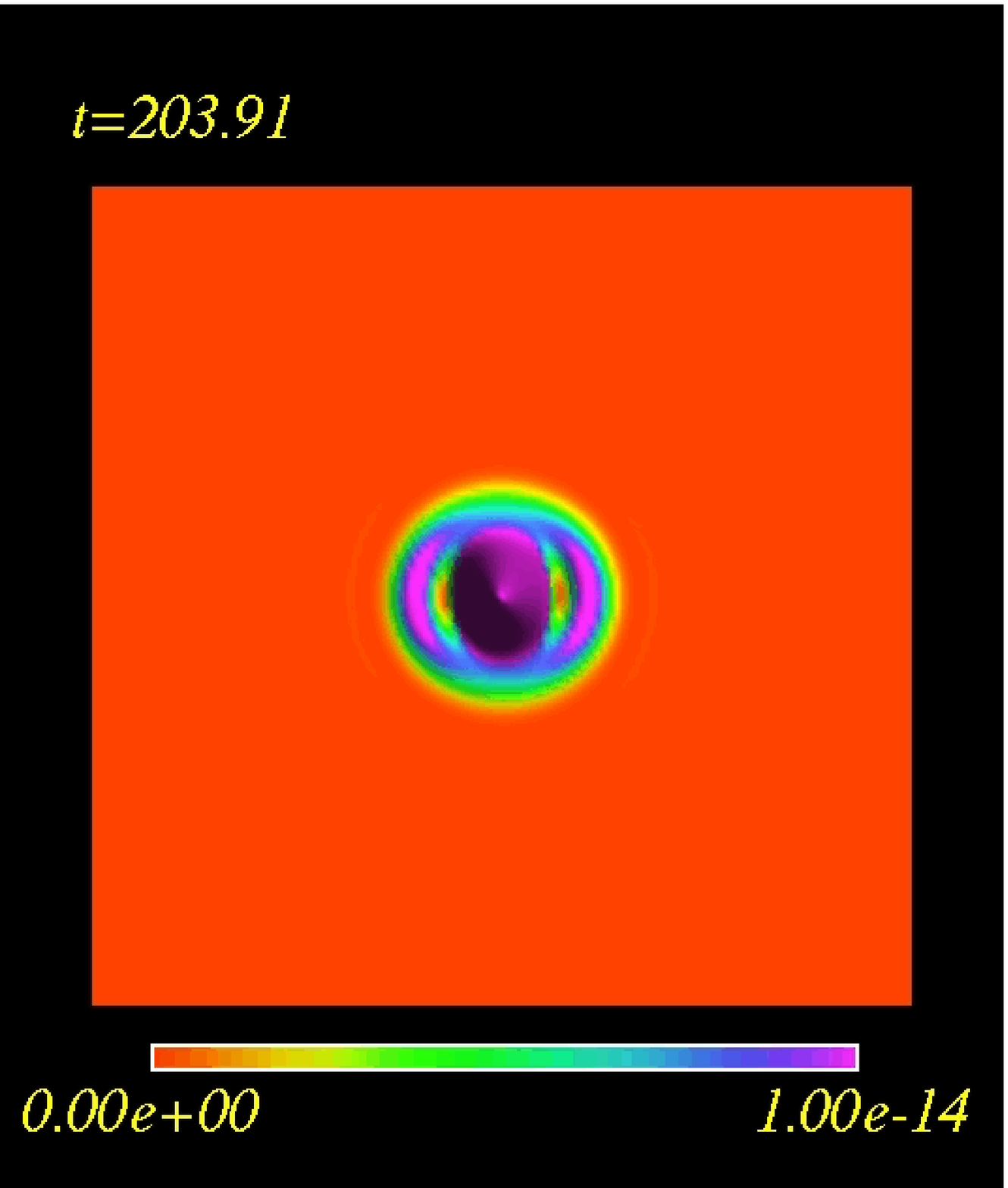,width=1.15in}}
  \caption{Evolution of a rotating solution (counter-clockwise) with magnetic moment inclined
   with respect to the $z$-axis of $\pi/4$ radians. Shown are the values of
  $|\vec{B}|^2$ on the $z=0$ plane for roughly half a period (from left to right). The computational domain extends in each direction a factor of four times the distance shown.
  }
  \label{fig:unaligned}
\end{center}
\end{figure}

Another interesting aspect to the dynamics of magnetized stars concerns
the behavior of the magnetic field. In particular, for the case in which
the magnetic moment of the star is not aligned with the rotational axis,
one can look at the structure of the dynamic magnetic field in terms of
poloidal and toroidal components.
In Fig.~\ref{fig:unaligned}, the evolution of rotating star with
a magnetic moment rotated away from the $z$-axis is shown. Because the initial
magnetic field is confined to the interior of the star, problems that generally
occur when the magnetic field is strong compared to the density
are avoided.

%
%
\noindent{\bf Acknowledgments:~~}
I thank J. Novak for his assistance with \magstar\ and 
Matthew Choptuik and Scott Noble for valuable discussions.
This work was supported by the National Science Foundation under grant
PHY-0325224 and
also 
through TeraGrid resources provided by SDSC under allocation award PHY-040027.

%
%
\bibliographystyle{ws-procs975x65}
\bibliography{main}

%
%
\end{document}